\documentclass[10pt,twocolumn,letterpaper]{article}

\usepackage{cvpr}              
\usepackage{graphicx}
\definecolor{cvprblue}{rgb}{0.21,0.49,0.74}
\usepackage[pagebackref,breaklinks,colorlinks,allcolors=cvprblue]{hyperref}

\def\paperID{20} 
\def\confName{PHAROSAIFMIH}
\def\confYear{2026}

\title{Attention Consistent Longitudinal Medical Visual Question Answering Guided by Vision Foundation Models}

\author{%
  Jialin Wu\\
  University of California, San Diego\\
  San Diego, CA 92037 \\
  {\tt\small jlwu@ucsd.edu} \\
  \and
  Qianru Zhang, Georges El Fakhri, Xiaofeng Liu \\
  Yale Biomedical Imaging Institute \\
  New Haven, CT 06510 \\
  {\tt\small xiaofeng.liu@yale.edu} \\
}

\begin{document}
\maketitle

\begin{abstract}
Longitudinal medical visual question answering (VQA) requires reasoning about anatomical differences between an image of a current time point and an image of a referred time point. We propose an attention-guided encoder-decoder for this task with chest X-rays. Instead of conventional direct contrast, we propose to include a lightweight affine registration module to reduce nuisance motion by co-registering the current image to the reference image with a small registration regularizer. The registered image pair is fed into the image encoder, followed by a frozen DINO-based mask generator and a trainable adaptive mask generator to produce masks applied to the original image pairs. The masked image pairs are again fed into the image encoder and concatenated with text features as the input to a multimodal transformer-based decoder to generate final answers. To facilitate learning stabilization and clarify the change signal, inspired by DINO-v3, we include additional auxiliary objectives, including a mask rebuilding loss, a pairwise Gram-style consistency loss, and a KoLeo uniformity loss, which enhances the geometry of the representation. On the Medical-Diff-VQA benchmark, the model delivers strong BLEU, ROUGE-L, CIDEr, and METEOR scores while offering intrinsic interpretability through the shared saliency mask. These results support saliency-conditioned generation with mild pre-alignment as a principled framework for longitudinal reasoning in medical VQA. Our training strategy also illustrates the potential of a paradigm in utilizing image foundation models in biomedicine: optimizing both supervised and unsupervised learning objectives simultaneously.
\end{abstract}    
\section{Introduction}
\label{sec:intro}

Medical Visual Question Answering (VQA) aims to answer open-ended clinical questions based on medical images, serving as a critical bridge from visual perception to clinical decision support ~\citep{survey}. In recent years, many medical VQA approaches have relied on pretrained visual or multimodal models~\citep{GLoRIA, llava, pmcvqa}. However, most of these works focus on a single time point that follows the natural image VQA tasks, whereas radiologists routinely need to compare current and previous studies to localize change, judge progression, and reconcile apparent discrepancies.

Longitudinal visual question answering (Diff-VQA) operationalizes this workflow by conditioning answers on paired images acquired at two time points, where the difference is often the signal of interest rather than the absolute appearance \citep{ekaid}. Recent benchmarks and methods for longitudinal chest X-rays have made this task concrete by supplying paired images, questions, and change-focused answers~\citep{mimic-diff-2, dataset1, dataset2}. Building on these resources, several approaches adapt vision–language models or design task-specific architectures to better capture temporal discrepancies, including prior work that emphasizes longitudinal pretraining \citep{plural}, residual alignment in the feature or pixel space \citep{reai}, or region-level retrieval and mixing \citep{regio}.
However, their attention at different time points is not explicitly encouraged to be consistent. Moreover, current approaches mainly focus on supervised fine-tuning, and the potential of introducing unsupervised objectives is not explored yet. Furthermore, they also suffer from the untransparency caused by the black box property of deep learning, which may cause disbelief and concerns from related stakeholders.
 
Saliency maps are a type of saliency visualization used to interpret deep learning models. In medical imaging tasks, they are widely employed to present verifiable evidence to clinicians and enhance model interpretability and trustworthiness~\citep{sal-driven, survey}. However, existing medical VQA models often treat saliency as a post-hoc explanation~\citep{medsal1:jin2021mapdoesfitall, medsal2, medsal3} rather than incorporating it as intrinsic supervision during training. In longitudinal settings, this is a missed opportunity because a consistent focus on corresponding anatomy across the two time points is essential to answering difference-type questions faithfully.

\begin{figure*}[t!]
  \centering
 \includegraphics[width=\linewidth]{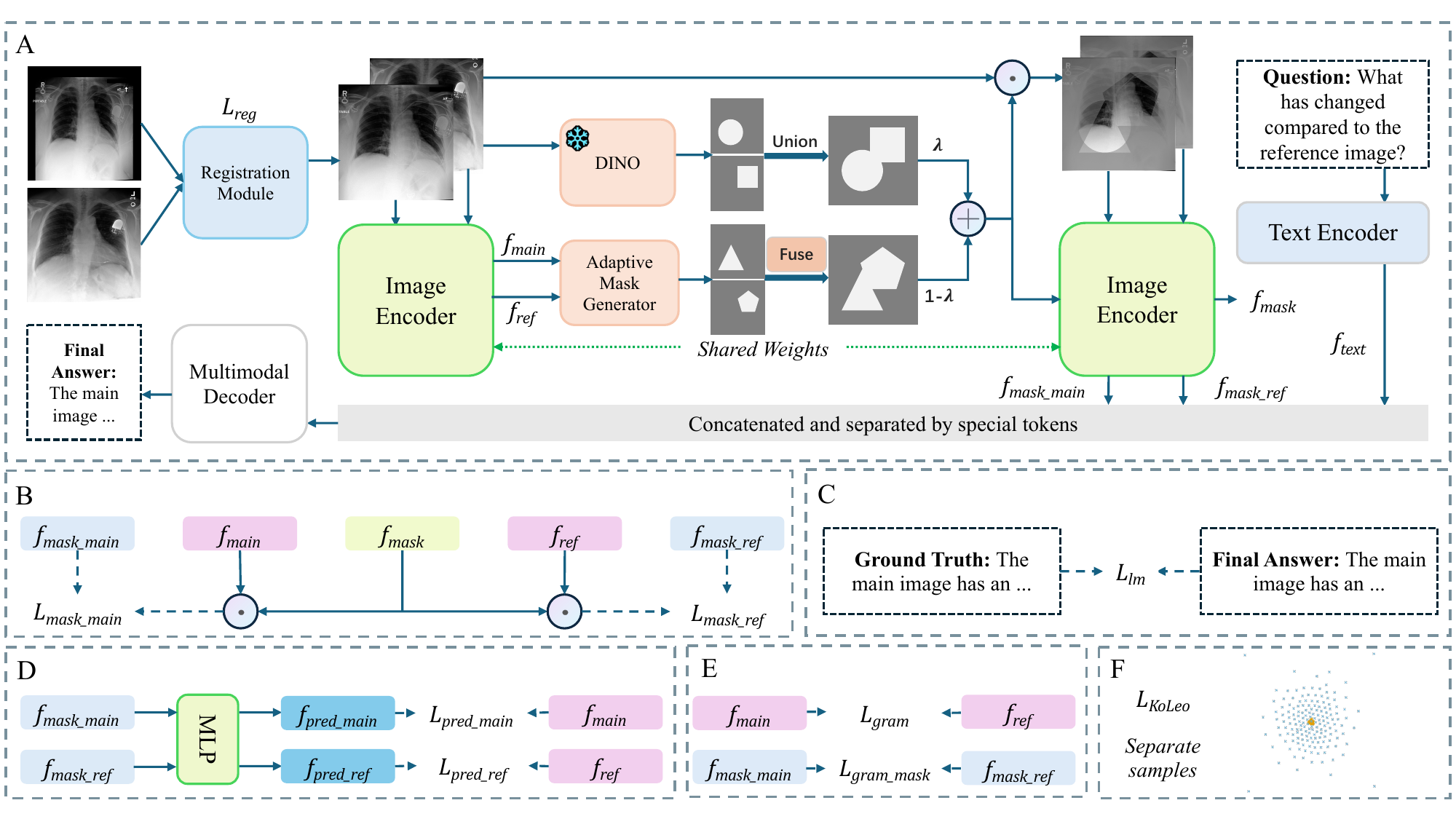}
  \caption{Illustration for our longitudinal medical VQA framework. \textbf{A.} First, perform approximately identical affine pre-registration (with parameter regularization $L_{\text{reg}}$) on the main and reference images, then extract features via the image encoder. The registered images are fed into DINO and an adaptive mask generator, respectively; both generate temporal masks under shared weights, which are then fused via the union/fuse function and weighted averaging (weights $\lambda$ and $1-\lambda$) to produce a change-aware shared mask applied uniformly to both registered images. The masked images are re-encoded, and the resulting two-time-point features are concatenated sequentially with the problem encoding using special separator tokens to form a multimodal prefix. Finally, the generative decoder outputs the answer. \textbf{B.} Mask consistency losses $L_{\text{mask\_main}}$ and $L_{\text{mask\_ref}}$ constrain the distance between the masked images' features and the product of the mask feature and images' features. \textbf{C.} Language modelling loss $L_{\text{lm}}$ directly supervises answer generation. \textbf{D.} Lightweight MLP prediction heads provide auxiliary supervision via $L_{\text{pred\_main}}$ and $L_{\text{pred\_ref}}$ respectively enhancing the semantic meaning of the masks. \textbf{E.} Gram-style similarity constraints are applied to both full and masked features ($L_{\text{gram}}$ and $L_{\text{gram\_mask}}$) to enforce similar patch structure between the main image and the reference image. \textbf{F.} KoLeo regularization $L_{\text{KoLeo}}$ promotes representational separability across samples. }
  \label{fig:arch}
\end{figure*}

To mitigate the above gaps, we introduce an attention-guided generative framework specifically designed for chest radiograph temporal comparison (\Cref{fig:arch}). The method has two design principles: (i) make the two images geometrically comparable and (ii) ensure that what the model says it cares about also determines where it looks at both time points, inspired by natural image co-attention~\citep{co-atten,co-sal}. Specifically, we have these modules. 
$\bullet$ \textbf{Micro pre-alignment.} A lightweight CNN-based module applies a near-identity affine warp to the current study to mitigate small pose and scale variations without overfitting or erasing true changes \citep{stn}. 
$\bullet$ \textbf{Dual-path mask generation.} One path employs the self-supervised visual prior DINO model~\citep{raddino} to provide robust lesion candidates, while the other path uses an adaptive mask generator to produce sample-adaptive masks from encoder features, with a hyperparameter changed during the training process, $\lambda$ controlling their relative proportions. $\bullet$ \textbf{Multigranularity training objectives;} Language modeling loss $L_{\text{lm}}$ optimizes answer quality; Mask consistency loss $L_{\text{mask\_main/ref}}$ constrains the difference between the inner products (of image features and the mask feature) and masked image features; a light-head prediction loss $L_{\text{pred\_main/ref}}$ enables mask rebuilding; Gram-style encourages similar patch-to-patch relationships and spatial structure similarity between images at different visits; and distribution normalization regularization $L_{\text{KoLeo}}$ enhances representational separability between samples and open-set robustness. 

The main contributions can be summarized as:

$\bullet$ We formalize a simple and effective way to enforce \emph{spatially consistent attention} across paired images by using a shared attention mask as a training signal for Diff-VQA. Specifically, we propose a plug-and-play mask generator that combines DINO priors with adaptive feature-driven masks without requiring additional annotations, balancing stability and sample adaptability.

$\bullet$ We use a comprehensive set of training objectives that cover classification and language semantics, representation alignments, spatial alignments, and attention alignments.

$\bullet$ We demonstrate competitive performance and incorporate interpretability and explainability in our framework, without the need for post-hoc saliency analysis, providing both textual answers and visual analysis of lesions. It alleviates the cognitive load on medical practitioners while mitigating distrust stemming from the black-box nature of deep learning models, demonstrating significant practical value.

\section{Related works}
\label{sec:rela}

\textit{Difference-aware medical VQA.} Medical-Diff-VQA~\citep{ekaid,mimic-diff-2} provides a large-scale benchmark of paired chest radiographs and has become a major evaluation dataset. Methodologically, early approaches often employed transfer learning from general image difference description (IDC) models as a strong baseline: MCCFormers~\citep{mccformers} utilized a transformer encoder-decoder architecture, performing multi-head attention similarity comparisons on patches from the two images. IDCPCL~\citep{idcpcl} aligns visual differences and text through self-supervised pre-training and contrastive learning, mitigating label scarcity. For medical applications, EKAID~\citep{ekaid} pioneered a systematic approach to differential Med-VQA, introducing graph-based representations based on expert knowledge. Subsequent approaches advanced along multiple trajectories: RegioMix~\citep{regio} employs region-level retrieval augmentation to retrieve question-relevant image regions prior to generation. PLURAL~\citep{plural} adapted Diff-VQA using a visual-language model pre-trained in two stages: natural text-image to longitudinal chest radiographs; ReAl~\citep{reai} combines generative responses with residual input and residual alignment of characteristics to explicitly highlight differences between two time phases; VED~\citep{ved} introduced embeddings of image differentiation, learning a distinct d-dimensional vector for each main/reference image and applying them to all visual tokens, allowing cross-attention decoding to distinguish images throughout the pipeline. 
 
\textit{Saliency and segmentation in medical images.} Research in this area demonstrates a converging trend from interpretable visualization towards spatial supervision and unified foundational models. Within chest radiography scenarios, a system benchmark~\citep{saportaBenchmarkingSaliencyMethods2022} reveals that multiple saliency methods (including Grad-CAM~\citep{grad-cam}) exhibit limited accuracy and stability in lesion localization.~\citet{attention_sal} proposed an attention-based transformer method for saliency generation in models for pneumothorax classification. Regarding segmentation models, nnU-Net~\citep{isenseeNnUNetSelfconfiguringMethod2021} provides a robust baseline for multimodal tasks through its self-configuration process. Subsequent approaches integrating or replacing U-Net with transformers (TransUNet~\citep{chen2021transunettransformersmakestrong}, Swin-UNet~\citep{cao2021swinunetunetlikepuretransformer}, UNETR~\citep{hatamizadeh2021unetrtransformers3dmedical}) further enhance global dependencies and multi-scale modelling. Concurrently, general segmentation foundational models are rapidly entering medical applications: MedSAM~\citep{medsam} demonstrates zero/few-shot generalization on million-scale medical datasets, while SAM2~\citep{ravi2024sam2} and MedSAM2~\citep{medsam2} extend promptable segmentation to 2D/3D and video domains. Regarding the representation backbone, DINO-based models are becoming a solid foundation for medical segmentation and saliency.
 
\textit{DINO backbones in radiology.}  DINOv2~\citep{dinov2} has been employed for training-free deformable medical image registration (DINO-Reg), securing first place in the OncoReg challenge~\citep{song2024generalpurposeimageencoder}. This demonstrates that semantic knowledge learned from natural images generalizes to cross-organ geometric alignment scenarios in medical data. For single-modality data like chest radiographs, RAD-DINO demonstrates strong competitiveness across classification/segmentation and image-text alignment tasks~\citep{raddino}. For multimodal scenarios such as MRI, MM-DINOv2 introduces multimodal patch embedding and whole-modality masking DINOv2~\citep{SchDan_MMDINOv2_MICCAI2025}.
Furthermore, interpretability efforts integrating DINOv2 representations (e.g., combining ViT-CX causal explanations with self-supervised features) provide evidence for clinical traceability~\citep{hussienExplainableSelfsupervisedLearning2025}.
Building upon DINOv3~\citep{dinov3}, SegDINO~\citep{yang2025segdinoefficientdesignmedical} achieves strong competitiveness across multiple medical segmentation benchmarks using a frozen DINOv3 with a lightweight decoder paradigm, while MedDINOv3~\citep{li2025meddinov3adaptvisionfoundation} attains or surpasses SOTA on segmentation tasks through multi-scale token aggregation and domain-adaptive pre-training on 3.87 million CT slices. In general, DINOv2 provides robust general representations and cross-task transferability, while DINOv3 further enhances high-resolution medical segmentation.

\section{Methods}
\label{sec:meth}

The pipeline has two components,  including a micro image registration module and a keyword-conditioned saliency extraction module, followed by image–text encoders and a multimodal decoder.

\subsection{Micro image registration module}
Given a main image \(I_{\text{main}}\in\mathbb{R}^{3\times H\times W}\) and a reference image \(I_{\text{ref}}\in\mathbb{R}^{3\times H\times W}\), a shallow CNN predicts 2D affine parameters \(\Theta=[A\;\mathbf{t}]\in\mathbb{R}^{2\times 3}\). We warp only the main image with a differentiable grid sampler:
\begin{equation}
\mathbf{x} = A\,\mathbf{x}_{\mathrm{tgt}} + \mathbf{t}.
  \label{eq:warp}
\end{equation}

To keep the transform near identity and avoid erasing true anatomical change, we regularize
\begin{equation}
\mathcal{L}_{\text{reg}} = w_{\text{sml}}\|\Theta - I\|_F^{2} + w_{\det}(\det(A)-1)^{2} + w_{\text{tran}}\|\mathbf{t}\|_2^{2}
  \label{eq:regloss}
\end{equation}
with \(w_{\text{sml}}=10^{-4}\), \(w_{\det}=10^{-5}\), and \(w_{\text{tran}}=10^{-6}\). The registered image is \(\widehat{I}_{\text{main}}\).

\subsection{Mask generating module}\label{subsec:mask}
Let $\phi(\cdot)$ be the shared image encoder and projector. We first extract the original image features:
\begin{equation}
f_{\text{main}}=\phi(\widehat{I}_{\text{main}})\in\mathbb{R}^{N\times C},\quad f_{\text{ref}}=\phi(I_{\text{ref}})\in\mathbb{R}^{N\times C}.
\label{eq:premask}
\end{equation}

From a frozen DINO branch, we obtain attention maps and a union prior as shown in~\Cref{eq:dino_union}. We use the cosine similarity between the embeddings of the CLS token and the patch token to represent the models' attention on that specific patch and then reshape the spatial format.
\begin{equation}
A_{\text{main}},A_{\text{ref}}\in[0,1]^{H\times W},\quad U=\max\!\left(A_{\text{main}},A_{\text{ref}}\right).
\label{eq:dino_union}
\end{equation}

A lightweight 3-layer MLP head $g(\cdot)$ produces per-token mask probabilities, and a fusion module $h(\cdot)$ using a 1-layer CNN gated head to fuse the union, intersection, and difference of these two masks' features.

\begin{equation}\label{eq:mask_head}
m_{\text{main}}=\sigma\!\big(g(f_{\text{main}})\big), \quad m_{\text{ref}}=\sigma\!\big(g(f_{\text{ref}})\big),
\end{equation}
\begin{equation}\label{eq:mask_fuse}
F=\sigma\!\big(h(m_{\text{main}}, m_{\text{ref}})\big).
\end{equation}

After reshaping the tokens into the image grid, the final mask combines the prior and adaptive components. In the experiment, we set $\lambda = 1$ for the initial and $\lambda = 0.5$ for the end. We use the cosine function for the intermediate value change.
\begin{equation}\label{eq:final_mask}
M=\lambda\,U+(1-\lambda)\,F ,\quad \lambda\in[0,1]
\end{equation}
 
We apply $M$ to both visits and re-encode the masked images using the same image encoder:
\begin{equation}\label{eq:masked_imgs}
I'_{\text{main}}=M\odot \widehat{I}_{\text{main}},\quad I'_{\text{ref}}=M\odot I_{\text{ref}}.
\end{equation}
 
\begin{equation}\label{eq:fmask}
\begin{aligned}
f_{\text{mask\_main}} &= \phi\!\big(I'_{\text{main}}\big),\quad
f_{\text{mask}}       &= \phi\!\big(M\big) \\
f_{\text{mask\_ref}}  &= \phi\!\big(I'_{\text{ref}}\big) 
\end{aligned}
\end{equation}

\subsection{Image encoder and projector}
We use a Swin-base model (patch size of 4 × 4 and window size of 12 × 12)~\citep{swin} first pretrained on ImageNet-21k~\citep{imgnet-21k} and further trained on classification tasks on the MIMIC-CXR dataset with CheXpert labels~\citep{ved} as our image encoder backbone.
Its penultimate feature map, \(\mathbb{R}^{N\times C}\), directly provides  a token sequence compatible with GPT-2~\citep{gpt2}input. A projector with one linear layer, one 8-head transformer encoder~\citep{transformer}, and a two-layer MLP maps image tokens to the text-representational space.

\subsection{Text encoder}
Questions are tokenized with embeddings shared by the decoder, then added with a learnable positional embedding, and passed through 6 12-head Transformer encoder layers.

\subsection{Multimodal decoder}
A GPT-2 small~\citep{gpt2} decoder from HuggingFace~\citep{hf} consumes masked image tokens and question tokens to generate the answer. We add special tokens: \(\langle\text{pad}\rangle\), \(\langle\text{img}\rangle\), \(\langle\text{qtn}\rangle\), and \(\langle\text{ans}\rangle\). Denote the representation of the question as $f_{\text{text}}$, using $[]$ to represent the concatenation; the input sequence $C$ is
\begin{equation}
[\langle\text{img}\rangle, f_{\text{mask\_main}}, \langle\text{img}\rangle, f_{\text{mask\_ref}}, \langle\text{qtn}\rangle, f_{\text{text}}, \langle\text{ans}\rangle].
  \label{eq:concat}
\end{equation}

\subsection{Training}\label{subsec:train}
We optimize the decoder for answer quality and add four auxiliary terms to (i) tie the masked pass to its gated counterpart, (ii) keep each visit individually diagnosable and constrain the information loss from applying masks, (iii) enforce longitudinal consistency, and (iv) avoid representation collapse. All losses are summed with the hyperparameter weights described at the end of this subsection. $N$ denotes the number of samples in the training set.

\textit{Language modeling.} Teacher-forcing cross-entropy $\mathrm{CE}$ on ground-truth answers $y$:
\begin{equation}\label{eq:llm}
\mathcal{L}_{\text{lm}}=\mathrm{CE}\!\left(\mathrm{Decoder}(C),\,y\right).
\end{equation}

\textit{Mask consistency.} The masked responses should equal the original responses gated by the final mask \(M\):
\begin{equation}\label{eq:lmask_main_ref}
\begin{aligned}
\mathcal{L}_{\text{mask\_main}}&=\frac{1}{N}\sum_{i=1}^{N}\Big\|f_{\text{mask\_main},i}-M_i\,f_{\text{main},i}\Big\|_2^{2}.\\
\mathcal{L}_{\text{mask\_ref}}&=\frac{1}{N}\sum_{i=1}^{N}\Big\|f_{\text{mask\_ref},i}-M_i\,f_{\text{ref},i}\Big\|_2^{2}.
\end{aligned}
\end{equation}

\textit{Reconstruction of the light head mask} A tiny head \(P(\cdot)\) regresses the masked tokens back to their pre-mask counterparts to preserve diagnosticability:
 
\begin{equation}\label{eq:pred_heads}
f_{\text{pred\_main}}=P\!\big(f_{\text{mask\_main}}\big),\quad
f_{\text{pred\_ref}}=P\!\big(f_{\text{mask\_ref}}\big),
\end{equation}
 
\begin{equation}\label{eq:lpreds}
\begin{aligned}
\mathcal{L}_{\text{pred\_main}}&=\frac{1}{N}\sum_{i=1}^{N}\Big\|f_{\text{pred\_main},i}-f_{\text{main},i}\Big\|_2^{2},\\
\mathcal{L}_{\text{pred\_ref}}&=\frac{1}{N}\sum_{i=1}^{N}\Big\|f_{\text{pred\_ref},i}-f_{\text{ref},i}\Big\|_2^{2}.
\end{aligned}
\end{equation}

\textit{Gram-style longitudinal consistency.} Inspired by the gram anchoring adapted in ~\citet{dinov3}, we also consider a gram-style objective here. Specifically, rather than computing the gram loss between teacher and student models, we computed the gram loss between the main image and the reference image to preserve the patch-to-patch relationship across different visits.
 
\begin{equation}\label{eq:lgram}
\begin{aligned}
\mathcal{L}_{\text{gram}}&=\Big\|G(f_{\text{main}})-G(f_{\text{ref}})\Big\|_F^{2},\\
\mathcal{L}_{\text{gram,mask}}&=\Big\|G(f_{\text{mask\_main}})-G(f_{\text{mask\_ref}})\Big\|_F^{2},
\end{aligned}
\end{equation}
 
\begin{equation}
\label{eq:gram_row}
\operatorname{Gram}(X) \;=\; \frac{1}{N}\,\hat X \hat X^\top,\quad
\hat X \;=\; \frac{X}{\lVert X\rVert_2}\in\mathbb{R}^{N\times C}.
\end{equation}

\textit{KoLeo dispersion.} We penalize small nearest-neighbor distances within each batch $B$ to avoid sample collapse:
\begin{equation}\label{eq:lkoleo}
\begin{aligned}
\mathcal{L}_{\text{KoLeo}}&=-\frac{1}{B}\sum_{i=1}^{B}\log\!\Big(\min_{j\neq i}\big\|\hat{z}_i-\hat{z}_j\big\|_2+\varepsilon\Big),\\ \hat{z}_i&=\frac{z_i}{\|z_i\|_2}.
\end{aligned}
\end{equation}

\textit{Total objective.} We combine language modeling, registration, and auxiliaries to form the total loss. In our settings, we have $\alpha_{\text{mask}} = \alpha_{\text{pred}} = \alpha_{\text{gram}} = 0.1$, and $\alpha_{\text{kl}} = 0.001$.
 
\begin{equation}\label{eq:ltotal}
\begin{aligned}
\mathcal{L}_{\text{total}}
&= \mathcal{L}_{\text{lm}}
+\mathcal{L}_{\text{reg}}
+\alpha_{\text{mask}}\!\big(\mathcal{L}_{\text{mask\_main}}+\mathcal{L}_{\text{mask\_ref}}\big)\\
&\quad+\alpha_{\text{pred}}\!\big(\mathcal{L}_{\text{pred\_m}}+\mathcal{L}_{\text{pred\_r}}\big)\\
&\quad
+\alpha_{\text{gram}}\!\big(\mathcal{L}_{\text{gram}}+\mathcal{L}_{\text{gram\_mask}}\big)+\alpha_{\text{kl}}\mathcal{L}_{\text{KoLeo}}.
\end{aligned}
\end{equation}

\textit{Two stages of training.} To avoid interfering with or compromising existing semantic information within the pretrained image encoder during the initial stages of multi-task learning, we divided the training into two phases. In the first phase, we froze the image encoder's weights and trained the model for four epochs, allowing the remaining components to learn their respective functionalities. In the second phase, we unfroze the image encoder and trained the entire model for a further four epochs.

\subsection{Inference}
During inference, we use the entire learned framework to generate the final answers.

\subsection{Rationality analysis of masking}
\label{sec:mask_theory}
  
Notice that if $M\equiv\mathbf{1}$, the model is reduced to the baseline without masking. Therefore, the masked model strictly contains the unmasked model as a special case. Denote the question as $q$ and the answer as $y$.

\textit{Masking as sufficient statistics.}
For longitudinal questions, we assume that the answer depends only on changes inside a sparse anatomical support $R\subset\{1,\dots,H\}\times\{1,\dots,W\}$:
\begin{equation}
\label{eq:region-assumption}
p(y\,|\,\hat I_{\text{main}}, I_{\text{ref}}, q)
=
p\big(y\,\big|\hat I_{\text{main}}^{(R)}, I_{\text{ref}}^{(R)}, q\big),
\end{equation}
where $\hat I^{(R)}$ keeps only pixels in $R$.  
Define the ideal binary mask
\begin{equation}
M^\star_{ij} =
\begin{cases}
1, & (i,j)\in R,\\[2pt]
0, & \text{otherwise},
\end{cases}
\end{equation}
and let $I_{\text{main}}^{\prime\star}=M^\star\odot\hat I_{\text{main}}$ and
$I_{\text{ref}}^{\prime\star}=M^\star\odot I_{\text{ref}}$.  
Then $(I_{\text{main}}^{\prime\star}, I_{\text{ref}}^{\prime\star})$ is a sufficient statistic for $y$ given $(\hat I_{\text{main}}, I_{\text{ref}}, q)$:
\begin{equation}
p(y\,|\,\hat I_{\text{main}}, I_{\text{ref}}, q)
=
p\big(y\,\big| I_{\text{main}}^{\prime\star}, I_{\text{ref}}^{\prime\star}, q\big).
\end{equation}
Therefore, a Bayesian optimal predictor exists that relies solely on the masked image. The objective of learning the mask $M$ is to approximate $M^\star$, filtering out task-irrelevant background without compromising information related to $y$.

\textit{Signal-to-noise and longitudinal differences.}
Write each registered image as a signal plus background noise,
\begin{equation}
\hat I_{\text{main}} = s_{\text{main}} + n_{\text{main}},\qquad
I_{\text{ref}}      = s_{\text{ref}}  + n_{\text{ref}},
\end{equation}
where $(s_{\text{main}}, s_{\text{ref}})$ encodes disease-related variations within $R$, whilst $n_{\text{main}}, n_{\text{ref}}$ denotes instrument and background noise, respectively. 
For $M$ approaching $M^\star$, we have
\begin{equation}
I'_{\text{main}} = M\odot\hat I_{\text{main}}
\approx s_{\text{main}} + (M\odot n_{\text{main}}),
\end{equation}
The noise energy decreases approximately in proportion to the area ratio $|R|/|\Omega|$ (where $\Omega$ denotes the support set of the entire image). 
In high-dimensional models, both reduced effective input dimensions and improved signal-to-noise ratios mitigate overfitting risks. Consequently, the mask serves as a structured regularization term, provided that it does not erroneously exclude the lesion regions.

\textit{Shared masks for longitudinal differences.}
Diff-VQA answers are driven by longitudinal changes rather than absolute findings, so the relevant quantity is the difference
\begin{equation}
\Delta I = \hat I_{\text{main}} - I_{\text{ref}}.
\end{equation}
Applying the \emph{same} mask $M$ to both time points yields
\begin{equation}
\Delta I' = I'_{\text{main}} - I'_{\text{ref}}
          = M\odot(\hat I_{\text{main}} - I_{\text{ref}})
          = M\odot\Delta I.
\end{equation}
This can be viewed as a projection operator $P_M$ acting on the joint input  $(\hat I_{\text{main}}, I_{\text{ref}})$ :
\begin{equation}
P_M(\hat I_{\text{main}}, I_{\text{ref}})
=
(M\odot\hat I_{\text{main}},\, M\odot I_{\text{ref}}),
\end{equation}
which preserves voxel-wise differences within the masked anatomical coordinates while attenuating all others. With our registration module, this enforces an inductive bias that the model must compare the corresponding anatomy across time, rather than relying on arbitrary global statistics.
\textit{DINO-guided prior and adaptive refinement.}
In the implementation, $M$ is obtained by combining the unsupervised prior $U$ generated by DINO and the task-adaptive mask $F$. $U$ constrains the mask to semantically and anatomically plausible regions, whilst $F$ task-specializes the prior under the joint influence of $\mathcal{L}_{\text{lm}}$, mask consistency loss, Gram-style constraints, and KoLeo regularization. Overall, this may be viewed as a straightforward estimation in the mask space: the prior derives from DINO, while the posterior is refined by the longitudinal VQA objective, yielding a shared mask that is both anatomically plausible and tightly aligned with the differential response.

\section{Experiments}

We use the longitudinal chest radiograph Diff-VQA dataset, Medical-Diff-VQA~\citep{ekaid,mimic-diff-2}, which constructs samples from paired studies of the same subject at two time points together with a difference-focused question–answer pair. The dataset is derived from MIMIC-CXR~\citep{MIMIC-CXR} and MIMIC-CXR-JPG~\citep{mimic-cxr-jpg} and was obtained from PhysioNet~\citep{physionet}. All usage follows the PhysioNet credentialed-access license and de-identification guidelines. The corpus contains 164{,}223 samples, split into 131{,}556 for training, 16{,}278 for validation, and 16{,}389 for testing. Further experimental details can be found in the supplementary materials. Following are details for each component shown in in \Cref{fig:arch}.

In the data preparation stage, we resized the input images as three-channel 384 × 384 pixel images. To mitigate overfitting of limited training data, we adjust brightness, contrast, and sharpness to diversify the training data. All codes are mainly implemented in two widely used python libraries, including PyTorch (2.6.0+cu124) and Hugging Face Transformers (4.56.2). The image encoder is from Hugging Face (\textit{microsoft/swin-base-patch4-window12-384-in22k}) and was further trained on one NVIDIA A100-SXM4-80G GPU. The full model was trained on four NVIDIA A100-SXM4-80G GPUs. Batch size is set to 10 for training and 8 for validation. The optimizer is AdamW with learning rate $1.5 \times 10^{-4}$ and weight decay $0.05$. The DINO backbone used in mask generation is from RAD-DINO~\citep{raddino}. 


\subsection{Quantitative Comparisons}
We adopt common generation metrics in VQA, BLEU-1~\citep{papineni-etal-2002-bleu} (1-gram precision with a brevity penalty), METEOR~\citep{banerjee-lavie-2005-meteor} (stem matching with an emphasis on recall), ROUGE-L~\citep{lin-2004-rouge} (overlap and longest common subsequence), and CIDEr~\citep{vedantam2015ciderconsensusbasedimagedescription} (a TF-IDF based consensus metric) to evaluate different aspects such as surface-level matching, semantic alignment, and consistency with human references. To emphasize the medical keyword and semantic meaning, also considering the scale of the metrics, we use CIDEr to select the best model at the end of training.
\begin{table}
  \caption{Evaluation on four metrics on Medical-Diff-VQA comparing ours with others. }
  \label{tab:compare}
  \centering
  \resizebox{\columnwidth}{!}{%
  \begin{tabular}{lcccc}
    \toprule
    Methods & BLEU-1 & METEOR & ROUGE-L & CIDEr \\
    \midrule
    MCCFormers~\citep{mccformers}  & 0.214 & 0.319 & 0.340 & 0     \\
    IDCPCL~\citep{idcpcl,reai}      & 0.614 & 0.303 & 0.582 & 0.703 \\ 
    EKAID~\citep{ekaid,reai}    & 0.628 & 0.339 & 0.557 & 1.027 \\
    RegioMix~\citep{regio}    & 0.705 & 0.381 & 0.651 & 1.804 \\
    PLURAL~\citep{plural}    & 0.704 & 0.381 & 0.653 & 1.832 \\
    VED~\citep{ved}  & {0.716} & 0.389 & 0.670 & {2.119} \\
    \midrule
    Ours    & {0.747} & {0.700} & {0.703} & 2.011 \\
    \bottomrule
  \end{tabular}
  }
\end{table}

Our approach demonstrates complementary advantages over existing methods across multiple evaluation metrics on the Medical-Diff-VQA task, exhibiting particularly significant superiority at the semantic level. Specifically, our approach achieves the highest BLEU-1 score of 0.747 on first-order n-grams, substantially outperforming VED (0.716). This indicates more comprehensive coverage of critical medical entities and difference-relevant vocabulary by our model.  More notably, on the METEOR metric, which prioritizes semantic matching and disambiguation, our method achieves 0.700, substantially surpassing all comparators, demonstrating the model's distinct advantage in capturing clinically critical information and generating semantically equivalent descriptions. On ROUGE-L, which is more sensitive to overall syntactic structure and paragraph coherence, our method achieved 0.703, substantially outperforming VED (0.670) and other baselines. Regarding CIDEr metrics, our score of 2.011, while slightly below  VED (2.119), substantially surpasses methods such as RegioMix (1.804) and PLURAL (1.832). This indicates robust expressive power in generating clinically relevant answers that integrate medical context with temporal variations in images. Overall, our method substantially surpasses existing differential VQA models in semantic adequacy and clinical relevance. It maintains parity with current state-of-the-art baselines in lexical accuracy and overall linguistic quality while achieving leadership in key metrics.

\subsection{Example and Qualitative Visualization}
\begin{figure*}[t]
  \centering
 \includegraphics[width=1\linewidth]{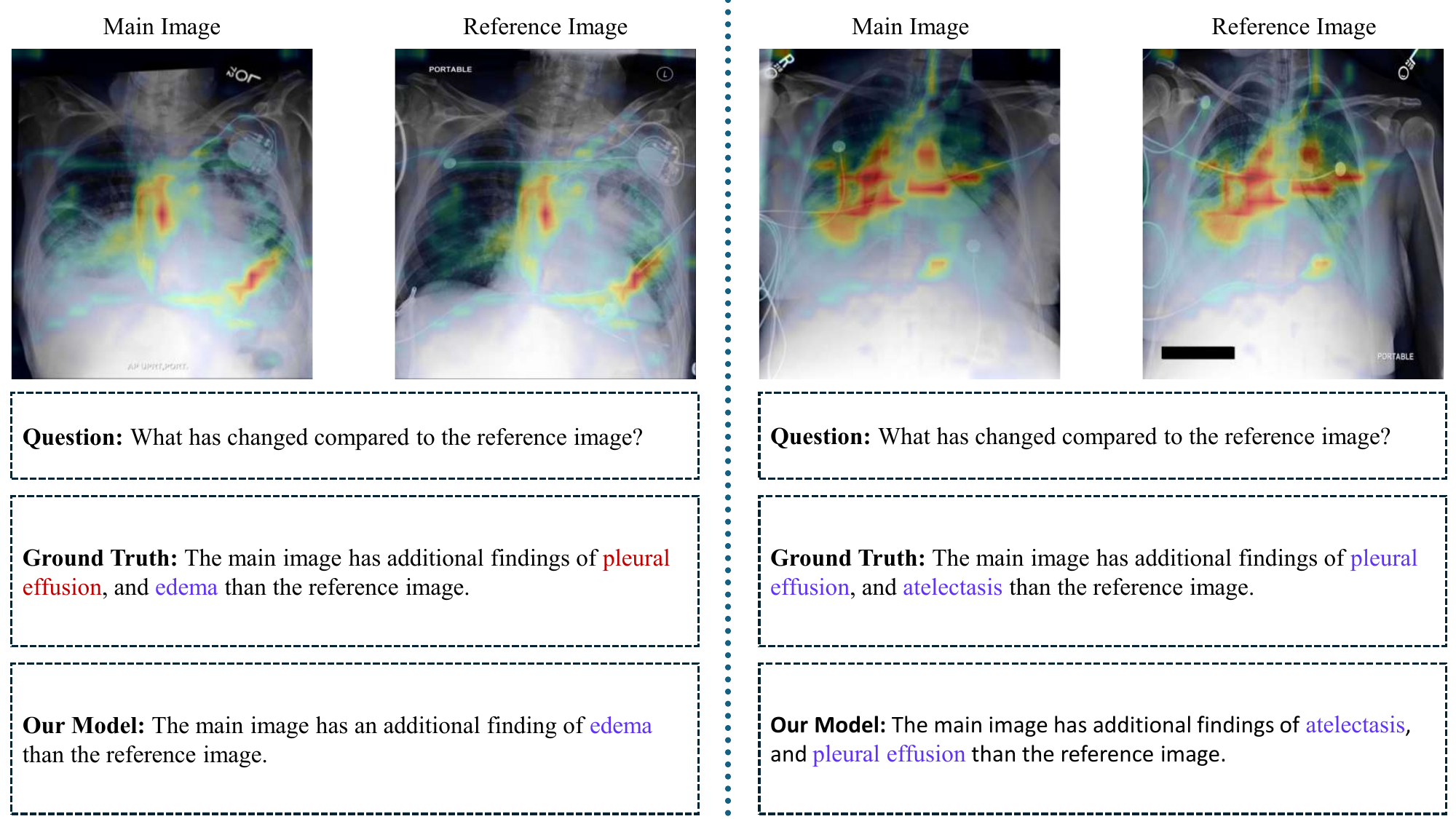}
  \caption{Examples of questions and their corresponding answers generated by our model and the ground truth. Correct predictions are highlighted in \textcolor{blue}{blue}, while incorrect predictions are highlighted in \textcolor{red}{red}.}
  \label{fig:example}
\end{figure*}

To further evaluate the actual effectiveness of the proposed method in practical application, we considered sampling from the test set to compare the genuine discrepancies between the model-generated answers and the ground truth. As shown in \Cref{fig:example}, although we achieved state-of-the-art performance on evaluation metrics, the generated answers could still differ from the actual answers. This aligns with findings in \citet{ved}, highlighting the need for new metrics tailored to the medical domain. A novel metric may require assigning greater weight to key medical terms and their semantic ordering, rather than treating them as ordinary tokens. Additionally, to further demonstrate how the proposed adaptive mask captures lesion regions, we overlaid the mask onto the original image. \Cref{fig:example} indicates our model's mask effectively identifies and focuses on critical areas in both images, underscoring the method's validity and exceptional interpretability. It is noteworthy that this artificial intelligence system is not intended to replace medical practitioners' diagnosis and treatment but rather to serve as an auxiliary tool for clinicians. Therefore, beyond providing textual prompts, this mask also offers reference and convenience for doctors interpreting images. The transparency it affords can enhance medical practitioners' trust in the model. It should be noted that some extramural points of interest remain visible beneath the mask, though these did not adversely affect the model's final performance. This suggests the model may utilize non-anatomical regions as a shortcut for inference. However, such covariates may exert differing influences on the model across other data distributions. Consequently, to enhance this model's generalizability and robustness, future research may explore further constraining and redistributing the model's attention using DINOv3. To further demonstrate the rationale and interpretability, additional examples of RAD-DINO and adaptive masks are provided in supplementary materials.

\subsection{Ablation studies}

In this section, we systematically analyze the impact of three key design choices on Diff-VQA performance through ablation experiments, with results presented in \Cref{tab:ablation}. Specifically, we compare four configurations: the full model incorporating all components; a model without the initial freezing of the image encoder; a model without the DINO-inspired unsupervised objectives; and a model without the saliency attention masks. Compared to the full model, performance metrics generally declined when the image encoder was not frozen earlier in training. This indicates that two-stage optimization helps stabilize visual representations, providing more discriminative features for subsequent modules and allowing sufficient time for them to learn their duties. Removing the DINO-inspired unsupervised objective resulted in a consistent decline in metrics such as BLEU, METEOR, and CIDEr scores, demonstrating that in the limited-annotation differential chest X-ray VQA scenario, additional unsupervised representation constraints help strengthen image-text alignment and enhance answer quality. Performance degraded markedly when inference was performed directly on the full image without applying the saliency attention mask. This demonstrates that saliency guidance is crucial for directing the model's focus towards lesions and regions of longitudinal change, thereby enabling the effective utilization of temporal variation information to generate accurate and reliable responses. Besides, this may also be partially caused by the fact that the mask consistency loss and mask rebuilding loss cannot be included in this scenario. Collectively, these results underscore the complementary roles and critical value of staged freezing, unsupervised objectives, and saliency attention masks in enhancing performance on Diff-VQA task.

\begin{table*}[htbp!]
  \caption{Results of ablation experiments to investigate different components' contribution.  $-$ refers to removing the component from the proposed framework. }
  \label{tab:ablation}
  \centering
  \resizebox{0.94\textwidth}{!}{%
  \begin{tabular}{llcccccccc}
    \toprule
     Model & BLEU-1 & BLEU-2 & BLEU-3 & BLEU-4 & METEOR & ROUGE-L & CIDEr \\
    \midrule
     Ours            & 0.747 & 0.620 & 0.510 & 0.425  & 0.700 & 0.703 & 2.011 \\
     $-$ Frozen Image Encoder for 4 epochs  & 0.711 & 0.581 & 0.473& 0.388 & 0.689 & 0.682  & 1.714 \\
     $-$ DINO-Inspired Unsupervised objectives & 0.699 & 0.576 & 0.471 & 0.390 & 0.690 & 0.671 & 1.765 \\
     $-$ Attention Masks       & 0.706 & 0.584 & 0.480 & 0.399 & 0.699 & 0.680 & 1.844 \\
    \bottomrule
  \end{tabular}
  }
\end{table*}

\section{Discussion and Conclusion}
\label{sec:conclusion}

This work proposes a generative framework for addressing longitudinal medical imaging differences, unifying approximate affine pre-registration, shared attention mask constraints, and multimodal generative decoding within an end-to-end system. First, a lightweight alignment is performed between the primary and reference examinations via the registration module, mitigating interference from patient position and acquisition conditions. Building upon this, a shared attention mask is generated using DINO priors and an adaptive feature-driven mask generator, simultaneously acting on both images. The Union/Fuse dual-channel architecture explicitly decouples common regions of interest and regions of change, enhancing sensitivity to lesion variations and spatial consistency without requiring additional pixel-level annotations; Subsequently, masked and unmasked image features undergo joint image encoding and projection. These are concatenated with question representations extracted by a text encoder to form a multimodal prefix. A generative decoder then outputs a differential descriptive answer, enabling traceable modelling of local lesion evolution and end-to-end response generation.

Nevertheless, this framework retains several limitations. Firstly, the introduction of DINO branches, adaptive mask generators, and multiple auxiliary losses inevitably increases computational overhead and implementation complexity during training. Further exploration is required to achieve a more optimal trade-off between efficiency and performance in resource-constrained scenarios. Secondly, the current registration module only supports approximately identical affine transformations. For cases with significant positional variations or markedly altered imaging conditions, its alignment capability may prove insufficient. Thirdly, as analyzed earlier, masks generated by the DINO branch still exhibit considerable noise. Future work could consider pre-training a segmentation head focused on the lesion. From an empirical perspective, we validated model performance on the MIMIC-CXR Dataset, without systematically evaluating generalization capabilities across other disease types, imaging modalities, or multi-center datasets.

Despite the aforementioned limitations, this study holds significant theoretical and practical implications: We present a straightforward yet effective implementation of explicitly constrained spatial attention on longitudinal chest radiograph pairs. By sharing masks and employing a dual-channel Union/Fuse architecture, we separately model anatomical regions requiring attention in both examinations and locally altered lesions, optimizing them collaboratively within a unified framework. This enables the mask generator to inherit the stable spatial prior provided by DINO while flexibly adjusting at the sample level through adaptive feature-driven mechanisms. Coupled with a comprehensive training objective that covers semantic modeling, representation alignment, and attention/feature distribution properties, the model achieves competitive question-answering performance while directly generating semantically consistent attention masks. This eliminates the need for additional post-processing significance analysis, providing clinicians with integrated textual responses and visual lesion highlights. Consequently, it reduces reading burden and mitigates distrust in black-box models. Lastly, it also provides a novel and principled approach to using existing image foundation models for biomedical research and applications.

\section{Acknowledgment}
This work is partially supported by the NIH grants P41EB022544, R21EB034911, and NVIDIA Academic Grant Program.

\clearpage
{
    \small
    \bibliographystyle{ieeenat_fullname}
    \bibliography{main}

\begin{thebibliography}{51}
\providecommand{\natexlab}[1]{#1}
\providecommand{\url}[1]{\texttt{#1}}
\expandafter\ifx\csname urlstyle\endcsname\relax
  \providecommand{\doi}[1]{doi: #1}\else
  \providecommand{\doi}{doi: \begingroup \urlstyle{rm}\Url}\fi

\bibitem[Banerjee and Lavie(2005)]{banerjee-lavie-2005-meteor}
Satanjeev Banerjee and Alon Lavie.
\newblock {METEOR}: An automatic metric for {MT} evaluation with improved correlation with human judgments.
\newblock In \emph{Proceedings of the {ACL} Workshop on Intrinsic and Extrinsic Evaluation Measures for Machine Translation and/or Summarization}, pages 65--72, Ann Arbor, Michigan, 2005. Association for Computational Linguistics.

\bibitem[Brima and Atemkeng()]{sal-driven}
Yusuf Brima and Marcellin Atemkeng.
\newblock Saliency-driven explainable deep learning in medical imaging: Bridging visual explainability and statistical quantitative analysis.
\newblock 17\penalty0 (1):\penalty0 18.

\bibitem[Cao et~al.(2021)Cao, Wang, Chen, Jiang, Zhang, Tian, and Wang]{cao2021swinunetunetlikepuretransformer}
Hu Cao, Yueyue Wang, Joy Chen, Dongsheng Jiang, Xiaopeng Zhang, Qi Tian, and Manning Wang.
\newblock Swin-unet: Unet-like pure transformer for medical image segmentation, 2021.

\bibitem[Chen et~al.(2021)Chen, Lu, Yu, Luo, Adeli, Wang, Lu, Yuille, and Zhou]{chen2021transunettransformersmakestrong}
Jieneng Chen, Yongyi Lu, Qihang Yu, Xiangde Luo, Ehsan Adeli, Yan Wang, Le Lu, Alan~L. Yuille, and Yuyin Zhou.
\newblock Transunet: Transformers make strong encoders for medical image segmentation, 2021.

\bibitem[Cho et~al.(2024)Cho, Kim, Shin, Cho, and Shin]{plural}
Yeongjae Cho, Taehee Kim, Heejun Shin, Sungzoon Cho, and Dongmyung Shin.
\newblock Pretraining vision-language model for difference visual question answering in longitudinal chest x-rays, 2024.

\bibitem[Gao et~al.(2020)Gao, Zhao, Liu, and Wang]{co-sal}
Guangshuai Gao, Wenting Zhao, Qingjie Liu, and Yunhong Wang.
\newblock Co-saliency detection with co-attention fully convolutional network, 2020.

\bibitem[Goldberger et~al.(2000)Goldberger, Amaral, Glass, Hausdorff, Ivanov, Mark, Mietus, Moody, Peng, and Stanley]{physionet}
Ary~L. Goldberger, Luis A.~N. Amaral, Leon Glass, Jeffrey~M. Hausdorff, Plamen~Ch. Ivanov, Roger~G. Mark, Joseph~E. Mietus, George~B. Moody, Chung-Kang Peng, and H.~Eugene Stanley.
\newblock Physiobank, physiotoolkit, and physionet.
\newblock \emph{Circulation}, 101\penalty0 (23):\penalty0 e215--e220, 2000.

\bibitem[Hatamizadeh et~al.(2021)Hatamizadeh, Tang, Nath, Yang, Myronenko, Landman, Roth, and Xu]{hatamizadeh2021unetrtransformers3dmedical}
Ali Hatamizadeh, Yucheng Tang, Vishwesh Nath, Dong Yang, Andriy Myronenko, Bennett Landman, Holger Roth, and Daguang Xu.
\newblock Unetr: Transformers for 3d medical image segmentation, 2021.

\bibitem[Hu et~al.()Hu, Gu, An, Zhang, {liu}, Kobayashi, Harada, Summers, and Zhu]{mimic-diff-2}
Xinyue Hu, Lin Gu, Qiyuan An, Mengliang Zhang, liangchen {liu}, Kazuma Kobayashi, Tatsuya Harada, Ronald Summers, and Yingying Zhu.
\newblock Medical-{{Diff-VQA}}: {{A Large-Scale Medical Dataset}} for {{Difference Visual Question Answering}} on {{Chest X-Ray Images}}.

\bibitem[Hu et~al.(2023)Hu, Gu, An, Zhang, Liu, Kobayashi, Harada, Summers, and Zhu]{ekaid}
Xinyue Hu, Lin Gu, Qiyuan An, Mengliang Zhang, Liangchen Liu, Kazuma Kobayashi, Tatsuya Harada, Ronald~M. Summers, and Yingying Zhu.
\newblock Expert knowledge-aware image difference graph representation learning for difference-aware medical visual question answering.
\newblock In \emph{Proceedings of the 29th ACM SIGKDD Conference on Knowledge Discovery and Data Mining}, page 4156–4165, New York, NY, USA, 2023. Association for Computing Machinery.

\bibitem[Huang et~al.(2021)Huang, Shen, Lungren, and Yeung]{GLoRIA}
Shih-Cheng Huang, Liyue Shen, Matthew~P. Lungren, and Serena Yeung.
\newblock Gloria: A multimodal global-local representation learning framework for label-efficient medical image recognition.
\newblock In \emph{2021 IEEE/CVF International Conference on Computer Vision (ICCV)}, pages 3922--3931, 2021.

\bibitem[Hussien et~al.()Hussien, Elkhateb, Saeed, Elsabawy, Elnakeeb, and Elrashidy]{hussienExplainableSelfsupervisedLearning2025}
Alaa Hussien, Abdelkareem Elkhateb, Mai Saeed, Nourhan~M. Elsabawy, Alaa~Ebraheem Elnakeeb, and Nora Elrashidy.
\newblock Explainable self-supervised learning for medical image diagnosis based on {{DINO V2}} model and semantic search.
\newblock 15\penalty0 (1):\penalty0 32174.

\bibitem[Isensee et~al.()Isensee, Jaeger, Kohl, Petersen, and Maier-Hein]{isenseeNnUNetSelfconfiguringMethod2021}
Fabian Isensee, Paul~F. Jaeger, Simon A.~A. Kohl, Jens Petersen, and Klaus~H. Maier-Hein.
\newblock {{nnU-Net}}: A self-configuring method for deep learning-based biomedical image segmentation.
\newblock 18\penalty0 (2):\penalty0 203--211.

\bibitem[Jaderberg et~al.(2016)Jaderberg, Simonyan, Zisserman, and Kavukcuoglu]{stn}
Max Jaderberg, Karen Simonyan, Andrew Zisserman, and Koray Kavukcuoglu.
\newblock Spatial transformer networks, 2016.

\bibitem[Jin et~al.(2021)Jin, Li, and Hamarneh]{medsal1:jin2021mapdoesfitall}
Weina Jin, Xiaoxiao Li, and Ghassan Hamarneh.
\newblock One map does not fit all: Evaluating saliency map explanation on multi-modal medical images, 2021.

\bibitem[Johnson et~al.()Johnson, Pollard, Berkowitz, Greenbaum, Lungren, Deng, Mark, and Horng]{MIMIC-CXR}
Alistair E.~W. Johnson, Tom~J. Pollard, Seth~J. Berkowitz, Nathaniel~R. Greenbaum, Matthew~P. Lungren, Chih-ying Deng, Roger~G. Mark, and Steven Horng.
\newblock {{MIMIC-CXR}}, a de-identified publicly available database of chest radiographs with free-text reports.
\newblock 6\penalty0 (1):\penalty0 317.

\bibitem[Johnson et~al.(2019)Johnson, Pollard, Greenbaum, Lungren, ying Deng, Peng, Lu, Mark, Berkowitz, and Horng]{mimic-cxr-jpg}
Alistair E.~W. Johnson, Tom~J. Pollard, Nathaniel~R. Greenbaum, Matthew~P. Lungren, Chih ying Deng, Yifan Peng, Zhiyong Lu, Roger~G. Mark, Seth~J. Berkowitz, and Steven Horng.
\newblock Mimic-cxr-jpg, a large publicly available database of labeled chest radiographs, 2019.

\bibitem[Lanfredi et~al.(2023)Lanfredi, Arora, Drew, Schroeder, and Tasdizen]{medsal2}
Ricardo~Bigolin Lanfredi, Ambuj Arora, Trafton Drew, Joyce~D. Schroeder, and Tolga Tasdizen.
\newblock Comparing radiologists' gaze and saliency maps generated by interpretability methods for chest x-rays, 2023.

\bibitem[Li et~al.(2023)Li, Wong, Zhang, Usuyama, Liu, Yang, Naumann, Poon, and Gao]{llava}
Chunyuan Li, Cliff Wong, Sheng Zhang, Naoto Usuyama, Haotian Liu, Jianwei Yang, Tristan Naumann, Hoifung Poon, and Jianfeng Gao.
\newblock Llava-med: Training a large language-and-vision assistant for biomedicine in one day, 2023.

\bibitem[Li et~al.(2025)Li, Wu, Lai, Hu, and Yang]{li2025meddinov3adaptvisionfoundation}
Yuheng Li, Yizhou Wu, Yuxiang Lai, Mingzhe Hu, and Xiaofeng Yang.
\newblock Meddinov3: How to adapt vision foundation models for medical image segmentation?, 2025.

\bibitem[Lin(2004)]{lin-2004-rouge}
Chin-Yew Lin.
\newblock {ROUGE}: A package for automatic evaluation of summaries.
\newblock In \emph{Text Summarization Branches Out}, pages 74--81, Barcelona, Spain, 2004. Association for Computational Linguistics.

\bibitem[Lin et~al.()Lin, Zhang, Tao, Shi, Haffari, Wu, He, and Ge]{survey}
Zhihong Lin, Donghao Zhang, Qingyi Tao, Danli Shi, Gholamreza Haffari, Qi Wu, Mingguang He, and Zongyuan Ge.
\newblock Medical visual question answering: {{A}} survey.
\newblock 143:\penalty0 102611.

\bibitem[Liu et~al.(2021)Liu, Lin, Cao, Hu, Wei, Zhang, Lin, and Guo]{swin}
Ze Liu, Yutong Lin, Yue Cao, Han Hu, Yixuan Wei, Zheng Zhang, Stephen Lin, and Baining Guo.
\newblock Swin transformer: Hierarchical vision transformer using shifted windows.
\newblock \emph{CoRR}, abs/2103.14030, 2021.

\bibitem[Lou et~al.(2025)Lou, Wang, Wu, Ng, White, Thakoor, Corcoran, Chen, and Liu]{medsal3}
Jianxun Lou, Huasheng Wang, Xinbo Wu, John Cho~Hui Ng, Richard White, Kaveri~A. Thakoor, Padraig Corcoran, Ying Chen, and Hantao Liu.
\newblock Chest x-ray visual saliency modeling: Eye-tracking dataset and saliency prediction model.
\newblock \emph{IEEE Transactions on Neural Networks and Learning Systems}, 36\penalty0 (9):\penalty0 16920--16930, 2025.

\bibitem[Lu et~al.(2024)Lu, Xie, Zeng, Lu, Wu, and Xia]{reai}
Zilin Lu, Yutong Xie, Qingjie Zeng, Mengkang Lu, Qi Wu, and Yong Xia.
\newblock { Spot the Difference: Difference Visual Question Answering with Residual Alignment }.
\newblock In \emph{proceedings of Medical Image Computing and Computer Assisted Intervention -- MICCAI 2024}. Springer Nature Switzerland, 2024.

\bibitem[Ma et~al.()Ma, He, Li, Han, You, and Wang]{medsam}
Jun Ma, Yuting He, Feifei Li, Lin Han, Chenyu You, and Bo Wang.
\newblock Segment anything in medical images.
\newblock 15\penalty0 (1):\penalty0 654.

\bibitem[Marhuenda et~al.(2025)Marhuenda, Obrador-Reina, Aas-Alas, Albiol, and Paredes]{ved}
Luis-Jesus Marhuenda, Miquel Obrador-Reina, Mohamed Aas-Alas, Alberto Albiol, and Roberto Paredes.
\newblock Unveiling differences: A vision encoder-decoder model for difference medical visual question answering.
\newblock In \emph{Medical Imaging with Deep Learning}, 2025.

\bibitem[Oh et~al.(2019)Oh, Kim, and Lee]{dataset1}
Dong~Yul Oh, Jihang Kim, and Kyong~Joon Lee.
\newblock Longitudinal change detection on chest x-rays using geometric correlation maps.
\newblock page 748–756, Berlin, Heidelberg, 2019. Springer-Verlag.

\bibitem[Oquab et~al.(2024)Oquab, Darcet, Moutakanni, Vo, Szafraniec, Khalidov, Fernandez, Haziza, Massa, El-Nouby, Assran, Ballas, Galuba, Howes, Huang, Li, Misra, Rabbat, Sharma, Synnaeve, Xu, Jegou, Mairal, Labatut, Joulin, and Bojanowski]{dinov2}
Maxime Oquab, Timothée Darcet, Théo Moutakanni, Huy Vo, Marc Szafraniec, Vasil Khalidov, Pierre Fernandez, Daniel Haziza, Francisco Massa, Alaaeldin El-Nouby, Mahmoud Assran, Nicolas Ballas, Wojciech Galuba, Russell Howes, Po-Yao Huang, Shang-Wen Li, Ishan Misra, Michael Rabbat, Vasu Sharma, Gabriel Synnaeve, Hu Xu, Hervé Jegou, Julien Mairal, Patrick Labatut, Armand Joulin, and Piotr Bojanowski.
\newblock Dinov2: Learning robust visual features without supervision, 2024.

\bibitem[Papineni et~al.(2002)Papineni, Roukos, Ward, and Zhu]{papineni-etal-2002-bleu}
Kishore Papineni, Salim Roukos, Todd Ward, and Wei-Jing Zhu.
\newblock {B}leu: a method for automatic evaluation of machine translation.
\newblock In \emph{Proceedings of the 40th Annual Meeting of the Association for Computational Linguistics}, pages 311--318, Philadelphia, Pennsylvania, USA, 2002. Association for Computational Linguistics.

\bibitem[P{\'e}rez-Garc{\'i}a et~al.(2025)P{\'e}rez-Garc{\'i}a, Sharma, Bond-Taylor, Bouzid, Salvatelli, Ilse, Bannur, Castro, Schwaighofer, Lungren, Wetscherek, Codella, Hyland, Alvarez-Valle, and Oktay]{raddino}
Fernando P{\'e}rez-Garc{\'i}a, Harshita Sharma, Sam Bond-Taylor, Kenza Bouzid, Valentina Salvatelli, Maximilian Ilse, Shruthi Bannur, Daniel~C. Castro, Anton Schwaighofer, Matthew~P. Lungren, Maria~Teodora Wetscherek, Noel Codella, Stephanie~L. Hyland, Javier Alvarez-Valle, and Ozan Oktay.
\newblock Exploring scalable medical image encoders beyond text supervision.
\newblock \emph{Nature Machine Intelligence}, 2025.

\bibitem[Radford et~al.(2019)Radford, Wu, Child, Luan, Amodei, and Sutskever]{gpt2}
Alec Radford, Jeff Wu, Rewon Child, David Luan, Dario Amodei, and Ilya Sutskever.
\newblock Language models are unsupervised multitask learners.
\newblock 2019.

\bibitem[Ravi et~al.(2024)Ravi, Gabeur, Hu, Hu, Ryali, Ma, Khedr, R{\"a}dle, Rolland, Gustafson, Mintun, Pan, Alwala, Carion, Wu, Girshick, Doll{\'a}r, and Feichtenhofer]{ravi2024sam2}
Nikhila Ravi, Valentin Gabeur, Yuan-Ting Hu, Ronghang Hu, Chaitanya Ryali, Tengyu Ma, Haitham Khedr, Roman R{\"a}dle, Chloe Rolland, Laura Gustafson, Eric Mintun, Junting Pan, Kalyan~Vasudev Alwala, Nicolas Carion, Chao-Yuan Wu, Ross Girshick, Piotr Doll{\'a}r, and Christoph Feichtenhofer.
\newblock Sam 2: Segment anything in images and videos.
\newblock \emph{arXiv preprint arXiv:2408.00714}, 2024.

\bibitem[Ridnik et~al.(2021)Ridnik, Ben-Baruch, Noy, and Zelnik]{imgnet-21k}
Tal Ridnik, Emanuel Ben-Baruch, Asaf Noy, and Lihi Zelnik.
\newblock Imagenet-21k pretraining for the masses.
\newblock In \emph{Proceedings of the Neural Information Processing Systems Track on Datasets and Benchmarks}, 2021.

\bibitem[Saporta et~al.()Saporta, Gui, Agrawal, Pareek, Truong, Nguyen, Ngo, Seekins, Blankenberg, Ng, Lungren, and Rajpurkar]{saportaBenchmarkingSaliencyMethods2022}
Adriel Saporta, Xiaotong Gui, Ashwin Agrawal, Anuj Pareek, Steven Q.~H. Truong, Chanh D.~T. Nguyen, Van-Doan Ngo, Jayne Seekins, Francis~G. Blankenberg, Andrew~Y. Ng, Matthew~P. Lungren, and Pranav Rajpurkar.
\newblock Benchmarking saliency methods for chest {{X-ray}} interpretation.
\newblock 4\penalty0 (10):\penalty0 867--878.

\bibitem[Scholz et~al.(2025)Scholz, Erdur, Ehm, Meyer-Baese, Peeken, Rueckert, and Wiestler]{SchDan_MMDINOv2_MICCAI2025}
Daniel Scholz, Ayhan~Can Erdur, Viktoria Ehm, Anke Meyer-Baese, Jan~C. Peeken, Daniel Rueckert, and Benedikt Wiestler.
\newblock { MM-DINOv2: Adapting Foundation Models for Multi-Modal Medical Image Analysis }.
\newblock In \emph{proceedings of Medical Image Computing and Computer Assisted Intervention -- MICCAI 2025}. Springer Nature Switzerland, 2025.

\bibitem[Selvaraju et~al.(2019)Selvaraju, Cogswell, Das, Vedantam, Parikh, and Batra]{grad-cam}
Ramprasaath~R. Selvaraju, Michael Cogswell, Abhishek Das, Ramakrishna Vedantam, Devi Parikh, and Dhruv Batra.
\newblock Grad-cam: Visual explanations from deep networks via gradient-based localization.
\newblock \emph{International Journal of Computer Vision}, 128\penalty0 (2):\penalty0 336–359, 2019.

\bibitem[Sim{\'e}oni et~al.(2025)Sim{\'e}oni, Vo, Seitzer, Baldassarre, Oquab, Jose, Khalidov, Szafraniec, Yi, Ramamonjisoa, Massa, Haziza, Wehrstedt, Wang, Darcet, Moutakanni, Sentana, Roberts, Vedaldi, Tolan, Brandt, Couprie, Mairal, J{\'e}gou, Labatut, and Bojanowski]{dinov3}
Oriane Sim{\'e}oni, Huy~V. Vo, Maximilian Seitzer, Federico Baldassarre, Maxime Oquab, Cijo Jose, Vasil Khalidov, Marc Szafraniec, Seungeun Yi, Micha{\"e}l Ramamonjisoa, Francisco Massa, Daniel Haziza, Luca Wehrstedt, Jianyuan Wang, Timoth{\'e}e Darcet, Th{\'e}o Moutakanni, Leonel Sentana, Claire Roberts, Andrea Vedaldi, Jamie Tolan, John Brandt, Camille Couprie, Julien Mairal, Herv{\'e} J{\'e}gou, Patrick Labatut, and Piotr Bojanowski.
\newblock {DINOv3}, 2025.

\bibitem[Song et~al.(2024)Song, Xu, and Yan]{song2024generalpurposeimageencoder}
Xinrui Song, Xuanang Xu, and Pingkun Yan.
\newblock General purpose image encoder dinov2 for medical image registration, 2024.

\bibitem[Vaswani et~al.(2023)Vaswani, Shazeer, Parmar, Uszkoreit, Jones, Gomez, Kaiser, and Polosukhin]{transformer}
Ashish Vaswani, Noam Shazeer, Niki Parmar, Jakob Uszkoreit, Llion Jones, Aidan~N. Gomez, Lukasz Kaiser, and Illia Polosukhin.
\newblock Attention is all you need, 2023.

\bibitem[Vedantam et~al.(2015)Vedantam, Zitnick, and Parikh]{vedantam2015ciderconsensusbasedimagedescription}
Ramakrishna Vedantam, C.~Lawrence Zitnick, and Devi Parikh.
\newblock Cider: Consensus-based image description evaluation, 2015.

\bibitem[Wang et~al.(2022)Wang, Qian, Fu, and Xue]{co-atten}
Qizao Wang, Xuelin Qian, Yanwei Fu, and Xiangyang Xue.
\newblock Co-attention aligned mutual cross-attention for cloth-changing person re-identification.
\newblock page 351–368, Berlin, Heidelberg, 2022. Springer-Verlag.

\bibitem[Wolf et~al.(2020)Wolf, Debut, Sanh, Chaumond, Delangue, Moi, Cistac, Rault, Louf, Funtowicz, Davison, Shleifer, von Platen, Ma, Jernite, Plu, Xu, Le~Scao, Gugger, Drame, Lhoest, and Rush]{hf}
Thomas Wolf, Lysandre Debut, Victor Sanh, Julien Chaumond, Clement Delangue, Anthony Moi, Pierric Cistac, Tim Rault, Remi Louf, Morgan Funtowicz, Joe Davison, Sam Shleifer, Patrick von Platen, Clara Ma, Yacine Jernite, Julien Plu, Canwen Xu, Teven Le~Scao, Sylvain Gugger, Mariama Drame, Quentin Lhoest, and Alexander Rush.
\newblock Transformers: State-of-the-art natural language processing.
\newblock In \emph{Proceedings of the 2020 Conference on Empirical Methods in Natural Language Processing: System Demonstrations}, pages 38--45, Online, 2020. Association for Computational Linguistics.

\bibitem[Wollek et~al.(2023)Wollek, Graf, {\v{C}}e{\v{c}}atka, Fink, Willem, Sabel, and Lasser]{attention_sal}
Alessandro Wollek, Robert Graf, Sa{\v{s}}a {\v{C}}e{\v{c}}atka, Nicola Fink, Theresa Willem, Bastian~O. Sabel, and Tobias Lasser.
\newblock Attention-based saliency maps improve interpretability of pneumothorax classification.
\newblock \emph{Radiology: Artificial Intelligence}, 5\penalty0 (2):\penalty0 e220187, 2023.
\newblock PMID: 37035429.

\bibitem[Yang et~al.(2025)Yang, Wang, Xing, Chen, and Zhu]{yang2025segdinoefficientdesignmedical}
Sicheng Yang, Hongqiu Wang, Zhaohu Xing, Sixiang Chen, and Lei Zhu.
\newblock Segdino: An efficient design for medical and natural image segmentation with dino-v3, 2025.

\bibitem[Yao et~al.(2018)Yao, Pan, Li, and Mei]{idcpcl}
Ting Yao, Yingwei Pan, Yehao Li, and Tao Mei.
\newblock Exploring visual relationship for image captioning, 2018.

\bibitem[Yue~Qiu and Satoh(2021)]{mccformers}
Kodai Nakashima Ryota Suzuki Kenji Iwata Hirokatsu~Kataoka Yue~Qiu, Shintaro~Yamamoto and Yutaka Satoh.
\newblock Describing and localizing multiple changes with transformers, 2021.

\bibitem[Yung et~al.(2024)Yung, Sivaraj, Stoyanov, Loukogeorgakis, and Mazomenos]{regio}
Ka-Wai Yung, Jayaram Sivaraj, Danail Stoyanov, Stavros Loukogeorgakis, and Evangelos~B. Mazomenos.
\newblock { Region-Specific Retrieval Augmentation for Longitudinal Visual Question Answering: A Mix-and-Match Paradigm }.
\newblock In \emph{proceedings of Medical Image Computing and Computer Assisted Intervention -- MICCAI 2024}. Springer Nature Switzerland, 2024.

\bibitem[Zambrano~Chaves et~al.()Zambrano~Chaves, Huang, Xu, Xu, Usuyama, Zhang, Wang, Xie, Khademi, Yang, Awadalla, Gong, Hu, Yang, Li, Gao, Gu, Wong, Wei, Naumann, Chen, Lungren, Chaudhari, Yeung-Levy, Langlotz, Wang, and Poon]{dataset2}
Juan~Manuel Zambrano~Chaves, Shih-Cheng Huang, Yanbo Xu, Hanwen Xu, Naoto Usuyama, Sheng Zhang, Fei Wang, Yujia Xie, Mahmoud Khademi, Ziyi Yang, Hany Awadalla, Julia Gong, Houdong Hu, Jianwei Yang, Chunyuan Li, Jianfeng Gao, Yu Gu, Cliff Wong, Mu Wei, Tristan Naumann, Muhao Chen, Matthew~P. Lungren, Akshay Chaudhari, Serena Yeung-Levy, Curtis~P. Langlotz, Sheng Wang, and Hoifung Poon.
\newblock A clinically accessible small multimodal radiology model and evaluation metric for chest {{X-ray}} findings.
\newblock 16\penalty0 (1):\penalty0 3108.

\bibitem[Zhang et~al.(2024)Zhang, Wu, Zhao, Lin, Zhang, Wang, and Xie]{pmcvqa}
Xiaoman Zhang, Chaoyi Wu, Ziheng Zhao, Weixiong Lin, Ya Zhang, Yanfeng Wang, and Weidi Xie.
\newblock {PMC-VQA}: Visual instruction tuning for medical visual question answering, 2024.

\bibitem[Zhu et~al.(2024)Zhu, Hamdi, Qi, Jin, and Wu]{medsam2}
Jiayuan Zhu, Abdullah Hamdi, Yunli Qi, Yueming Jin, and Junde Wu.
\newblock Medical sam 2: Segment medical images as video via segment anything model 2, 2024.

\end{thebibliography}
}



\end{document}